# A Computer Vision Approach To Identify Einstein Rings And Arcs

Chien-Hsiu Lee[1]*
[1]Subaru Telescope, National Astronomical Observatory of Japan
650 North Aohoku Place, Hilo, HI 96720, USA

**Abstract**
Einstein rings are rare gem of the strong lensing phenomena. Unlike doubly or quadrouply lensed systems which only provide lens mass profile constraints at 1 or 2 position angles, the ring images can be used to probe the underlying lens gravitational potential at every single position angles, thus enable us to put much tighter constraints on the lens mass profile. In addition, the magnified and stretched images of the background source also enable us to probe properties of high-z galaxies with enhanced spatial resolution and higher signal to noise ratios, which is otherwise not possible for un-lensed galaxy studies. Despite their usefulness, only a handful of Einstein rings have been repoted up-to-date, mainly based on serendipitous discoveries or visual inspections of hundred thousands of massive galaxies – i.e. luminous red galaxies (LRG) – or galaxy clusters – i.e. brightest cluster galaxy (BCG). In the era of large sky survey, and with the upcoming surveys such as Large Synoptic Survey Telescope, visual inspection to discovery Einstein rings is very difficult, if not impossible, and an automated approach to identify ring pattern in the big data to come is in high demand. Here we present an Einstein ring recognition approach based on computer vision techniques. The workhorse of this approach is the circle Hough transform, which can recognise circular patterns or arcs at any given position with any radius in the images. We devise a two-tier approach: first pre-select LRGs associated with multiple blue objects as possible lens galaxies, than feed these possible lenses to Hough transform to identify Einstein rings and arcs. As a proof-of-concept, we investigate our apprach using the Sloan Digital Sky Surveys. Our results show a 100% completeness, albeit a low purity at 40%. We also apply our approach to three newly discovered Einstein rings and arcs, in the Dark Energy Survey, Hyper Suprime-Cam Subaru Strategic Program, and UltraVISTA survey, illusting the versatiliy of our approach to on-going and up-coming large sky surveys in general. The beauty of our approach is that it is solely based on JPEG images, which can be easily obtained in batch mode from SDSS finding chart tools, without any pre-processing of the image. An implementation in Python will be available and can be downloaded from the author.

## 1 Introduction

According to Einstein's theory of general relativity (Einstein 1915), massive foreground galaxies or clusters of galaxies induce strong space-time curvature, serving as gravitational lenses and bend the lights of background sources to form multiple and magnified images along the observer's line-of-sight. Such strong lensing systems provide unique opportunities to study distant galaxies and cosmology in detail (see Treu 2010, for a recent review). On one hand, modeling the lensed images enables us to constrain the mass profile and dark matter distribution of the lens, as well as probing dwarf satellites associated with the lens galaxy, to test ΛCDM model(Vegetti et al. 2012). On the other hand, gravitational lensing in tandem with Hubble space telescope offers us a peek into galaxies at cosmic dawn (Zheng et al. 2012), beyond the reach of current instruments. In addition, joining the forces of supernova and gravitational lensing will provide insights to the expansion rate of our Universe (Refsdal 1964; Kelly et al. 2015).

When the foreground lens galaxy, the background source, and the observer are well aligned on a straight line, the lensed images will become a ring. Such phenomena were first pointed out by Chwolson (1924) and later remarked by Einstein (1936) upon the request of R. W. Mandl, hence bear the name Einstein-

*E-mail: leech@naoj.org

Chwolson ring (hereafter Einstein ring, for simplicity). Einstein rings are rare gems of the strong lensing phenomena; they provide tighter constraints on the lens than multiply-imaged strong lensing systems, because: 1) the enclosed lens mass can be directly measured from the angular size of the Einstein ring, without any prior assumptions on the lens mass profile and underlying lensing potential (see e.g. Narayan & Bartelmann 1996); 2) the lens modeling, in tandem with lens velocity dispersion gradient derived from spatially resolved spectra of integral field unit spectrograph observations, will provide a complete view of the mass distribution of baryonic and dark matter; 3) compared to doubly or quadruply lensed systems, the gravitational potential of the lens can be studied in great detail, because the ring structure can provide constraints at virtually every position angle (Kochanek, Keeton, & McLeod 2001); 4) the stretched and magnified source images offer us a unique opportunity to resolve the source with enhanced spatial resolution and higher signal to noise ratio, enabling us to study metallicity gradient across the source in detail (Jones et al. 2013).

Despite their usefulness, only dozens of partial Einstein ring systems have been discovered so far (see e.g. Bolton et al. 2008; Stark et al. 2013), mainly due to the rare chance to have lenses and sources well-aligned along the observer's line-of-sights. Among these partial Einstein ring systems, only a handful of them show complete or nearly complete ring morphology, e.g. B1938+666 (King et al. 1998), FOR J0332-2557 ($\sim 260^o$ Cabanac et al. 2005), Cosmic Horseshoe ($\sim 300^o$ Belokurov et al. 2007), Elliot Arc (Buckley-Geer et al. 2011), Canarias Einstein ring Bettinelli et al. (2016), Eye of Horus (Tanaka et al. 2016), to name a few. These Einstein rings are detected either serendipitously (e.g. Cabanac et al. 2005; Buckley-Geer et al. 2011; Bettinelli et al. 2016), or by examining images of massive galaxies as possible lens, containing tens of thousands of luminous red galaxies (LRGs) and brightest cluster galaxies (BCG) (e.g. Lin et al. 2009; Tanaka et al. 2016). With the advent of ultra-wide cameras and large area surveys, e.g. Sloan Digital Sky Survey[1] (SDSS, York et al. 2000), Pan-STARRS[2] (Kaiser et al. 2010), SkyMapper[3] (Schmidt et al. 2005), Dark Energy Survey[4] (DES, Flaugher et al. 2015), and Hyper Suprime-Cam Subaru Strategic Program[5] (HSC, Miyazaki et al. 2006), and Large Synoptic Survey Telescope[6] (LSST, LSST Science Collaboration et al. 2009), hundred thousands of lens are expected to be discovered (Collett 2015); it is very difficult, if not impossible, to visually inspect all the candidate LRG and BCG lenses in the big data to come[7].

In a more general context, there are several attempts to identify gravitational arcs or multiply lensed objects around massive galaxies, using either image-based or spectroscopy-based manners. For example, Belokurov et al. (2007) have shown that by simply selecting bright, red galaxies associated with multiple blue, faint objects in SDSS already yields several strong lens candidates with wide arcs. They further tailored the selecting criteria for SDSS, in order to eliminate false-positives from galaxy groups, mergers, and tidal tails, and presented more promising arc candidates in Belokurov et al. (2009). More sophisticated image-based approach involves high-resolution images with dedicated modelling. For example, Marshall et al. (2009); Gavazzi et al. (2014); Brault & Gavazzi (2015); Paraficz et al. (2016) have demonstrated the feasibility to remove the foreground lens light with dedicated modelling, and finding blue objects in the residual images, which are otherwise blended in the foreground lens light profile. On the other hand, searching for gravitational lensing systems by identifying higher redshift emission lines hiding in the spectra of foreground galaxies have been proposed (Warren et al. 1996; Hewett et al. 2000) and proved feasible (see e.g. Bolton et al. 2008; Smith 2016). Both detailed imaging and spectroscopic searches require extensive modelling and fitting. The color and brightness selection of Belokurov et al. (2007) is easier to implement, yet it may deliver false-positives outnumbered the Einstein rings or arcs if the selection criteria are set too loose and require heavy human interactions (Lin et al. 2009), or need further fine-tuning of the crietria (Belokurov et al. 2009) and cannot be applied to other surveys.

To make better use of the color and brightness selection method proposed by Belokurov et al. (2007) and to reduce human interactions during Einstein ring finding process, we propose to use a novel approach to identify circular patterns using computer vision techniques. We note that in this study we concentrate on recognition of circular patterns that are applicable to Einstein rings and arcs. The draw back is that such algorithm cannot identify potential quad lenses, which are in the form of elliptical patterns.

This paper is organized as follows. In Section 2 we present the idea of using computer vision to find circles and arcs. We then apply this approach to several ongoing wide-field surveys and present our results in Section 3, followed by prospects in Section 4.

---

[1] http://www.sdss.org
[2] http://pan-starrs.ifa.hawaii.edu
[3] http://www.mso.anu.edu.au/skymapper/
[4] https://www.darkenergysurvey.org
[5] http://hsc.mtk.nao.ac.jp/ssp/
[6] https://www.lsst.org

---

[7] We note that there are some progresses of visually classifying large data-set by involving citizen scientists, e.g. SPACE WARPS (More et al. 2016)

## 2 Method

Computer vision tackles tasks of reproducing the ability of human vision by electronically perceiving and understanding images, whose techniques can be easily and readily applied to astronomical data. One widely applied algorithm for computer vision to recognise parameterisable patterns is Hough transform (Hough 1962; Duda & Hart 1972), originally devised for identifying straight lines, but later generalised for any parameterisable patterns. The idea is, for a parameterisable shape, in our case a circle:

$$(x-a)^2 + (y-b)^2 = r^2, \quad (1)$$

the Hough transform constructs a quantized parameter space (dubbed "accumulator") for the circle's centre ($a$, $b$), and radius $r$. For every individual pixel on the image, the Hough transform then searches in the accumulator, and checks whether the pixels on the image can be described by a given set of parameters in the accumulator. This is considered as a voting process, and the set of parameters which obtain the highest numbers of votes, or pass a given threshold, will be regarded as a circle detection. For a $n \times n$ pixel image, Hough transform will construct a $n^3$ accumulator, hence it will be computational expensive to blindly search for circles in large sky images. However, the Einstein rings have certain advantages to reduce the computational cost, as we will discuss in the next section.

## 3 Applications

Hough transform has been used in astronomy for years, but mainly in removing satellite tracks (Cheselka 1999) or detecting echelle orders (Ballester 1994). It has only been recently recognised to use Hough transform to detect extended circular and/or arc-like astronomical objects, such as supernova remnants, bent-tailed galaxies, radio relics and so on, by Hollitt & Johnston-Hollitt (2012). However, despite the promising results of detecting circular / arc features, Hollitt & Johnston-Hollitt (2012) also pointed out that circle Hough transform will be computational expensive for blind search, and proposed a two tired approach. i.e. first find the location of possible extended circular objects, then perform Hough transform around this location, to reduce the computational efforts.

This being said, Einstein rings have several advantages to apply Hough transform. First of all, Einstein rings are mostly caused by massive objects, i.e. massive early type galaxies like luminous red galaxies, or galaxy clusters, which can be easily selected using their red colours. Secondly, the Einstein rings / arcs are relatively blue compared to the lens, hence a handy criteria to pre-select Einstein ring candidates would be red objects associated with multiple blue objects nearby.

Indeed, Belokurov et al. (2007) have shown that by selecting objects with $r < 19.5$ mag and $g - r > 0.6$ mag associated with multiple faint ($r > 19.5$) and blue ($g - r < 0.5$) sources within 6 arcseconds, they can find the Cosmic Horseshoe and the 8 o'clock arc. However, as Lin et al. (2009) showed, even employing the aforementioned criteria, there are still thousands of candidates remain for visual inspection, hence it is necessary to use circular pattern recognition algorithm as Hough transform mentioned here. Last but not the least, Einstein rings have relative small angular size ($\sim$ 5 - 10 arcseconds in diameter). Since we know the range of the radius parameter space *in priori*, we can greatly reduce the computation time for Hough transform.

As a proof-of-concept, we use SDSS images to test the robustness of detecting Einstein rings and arcs using circle Hough transform. Since the main problem of automatic lens finding is to clean out false-positives, e.g. ring glaxies, face-on spirals, tidal tails and chance alignments that appear like Einstein rings, we devise a two-tier process: 1) we pre-select LRGs associated with blue nearby objects; 2) we than feed these possible lens to circle Hough transform to identify Einstein rings and arcs.

To locate massive galaxies, We first start with the LRGs of Eisenstein et al. (2001), which can be easily identified using colour selections. We than search for multiple (n$\geq$2) blue ($g - r < 0.5$) faint ($r > 19.5$ mag) objects associated with the LRGs within 6 arcsecs. This reduce the number of possible lens LRGs from $\sim$ 200,000 to 1,624, which are further passed to Hough transform to identify circles and arcs.

We use the circle Hough transform implementation from OpenCV[8], an open source computer vision library. The program takes in an image of JPEG format to search for circular patterns. If the input is a color image, the program will convert the input into a 8-bit, single-channel, gray-scale image. For convenience, we obtain the SDSS color images from SDSS DR10 Finding Chart Tool[9] in batch mode, which can be tailored to retrieve images at a given position with different sizes and pixel scales. Since Einstein rings are relative small, we thus tailored our images to be 64×64 pixels, with pixel scale of 0.396 arcsec/pixel (the natural pixel scale of SDSS CCDs). We then feed the SDSS postage stamp images to the python version of OpenCV circle Hough transform. Since our images are rather small, the computational cost is negligible - it only takes 0.13 seconds to analyse each individual image on a 2.7 GHz Intel Core i5 processor. At the end the Hough transform reduce the number of possible LRGs lens from 1,624 to 10, where 4 of them are spectroscopically confirmed wide arc lenses, suggesting a 100% completeness and 40% purity of our approach. Despite the small size of the image, and the fast calculation time, the circle patterns of Einsteinrings are well identified, even if they are only partial arcs. The results of our analysis are shown in Fig. 1 and table 1.

---

[8] http://opencv.org
[9] http://skyserver.sdss.org/dr10/en/tools/chart/image.aspx

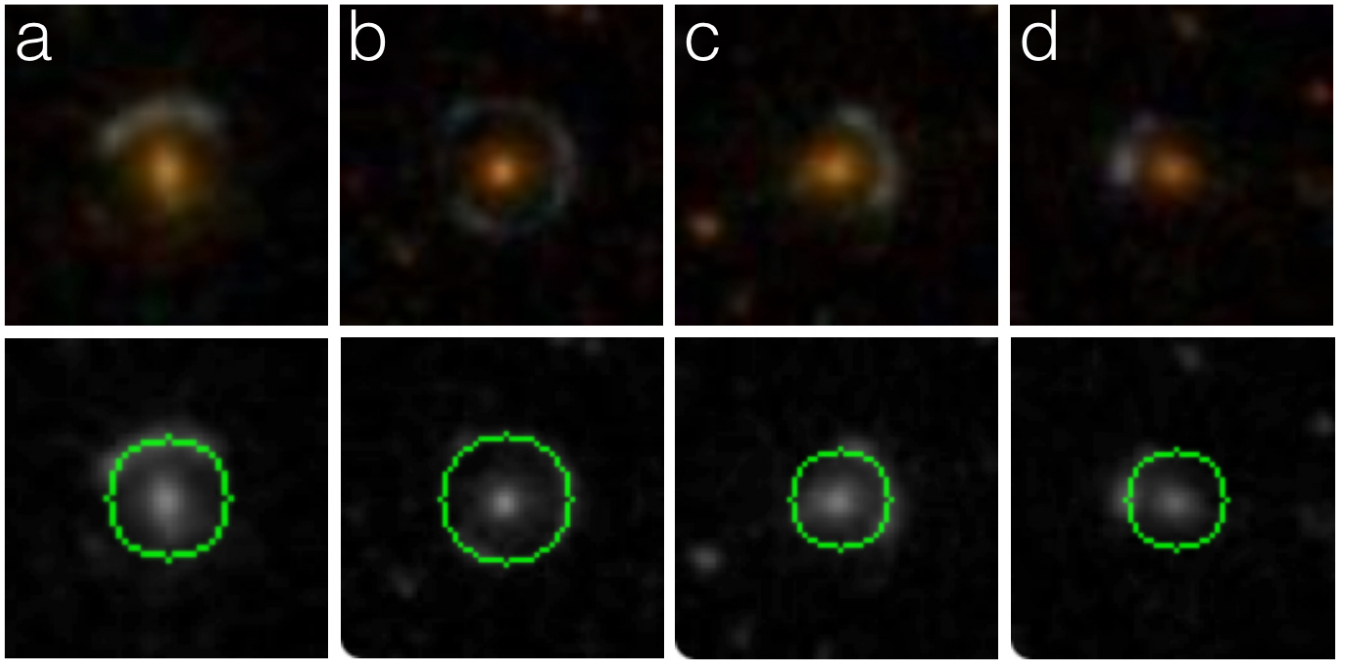

**Figure 1.** Applying Hough transform to Einstein rings from SDSS. Upper panels: images of a) 8 o'clock (Allam et al. 2007); b) Cosmic Horseshoe (Belokurov et al. 2007); c) Clone (Lin et al. 2009); d) CSWA 7 (Belokurov et al. 2009). Lower panels: Gray-scale images used for patterns recognition, and circular patterns identified by Hough transform (marked in green circles).

**Table 1** Selection criteria.

| Criterion | Number |
| --- | --- |
| Luminous red galaxies from SDSS | 200,367 |
| Multiple blue objects nearby | 1,624 |
| Hough transform | 10 |
| Known lens | 4 |

While we show some success with the SDSS data, the sample size of lens systems is very small. In addition, the SDSS examples are all composed of wide-separation arcs. In this regard, we also perform testing using a larger sample from the CFHT legacy data. We make use of the public available images from SPACE WARPS[10] (Marshall et al. 2016; More et al. 2016), which contains 59 high confident lens candidates identified by citizen scientists. Among the 59 SPACE WARPS lens systems, the circular Hough Transform is able to identify 53 lenses. The failed 6 cases are either arcs with small curvature or heavily blended lenses. In the first cases, because the arcs have very small curvature and resemble a line, the Hough transform failed to find a reasonable fit to a piece-wise circle/ring. In the later cases, because the arcs are heavily blended in the lens light, the Hough transform cannot distinguish the arcs from the lens light profile, thus failed to identify a circular pattern.

To test the versatility of our approach to different surveys, we also apply the Hough transform to Einstein rings discovered by DES (Canarias Einstein ring Bettinelli et al. 2016) and by HSC (Eye of Horus Tanaka et al. 2016), both are serendipitous discoveries. The color images are taken from the discovery papers. As shown in Fig. 2, the circle Hough transform can identify both rings in DES and HSC without any ambiguity, demonstrating the versatility of our approach. Furthermore, since the lens of Eye of Horus is a BCG, it also shows that the Hough transform can work in crowded fields in galaxy clusters.

In addition, we also test the Hough transform on a quadruply lensed quiescent galaxy, COSMOS 0050+4901 (Hill et al. 2016), discovered by the COSMOS/UltraVISTA survey. As the lensed images are relatively red, this demonstrates that the Hough transform can identify Einstein rings / arcs regardless of the source colours.

## 4 Prospects

We have demonstrated the efficiency and robustness of finding Einstein rings, even partial arcs, in the on-going large area surveys using the circle Hough transform. Though we only demonstrate its feasibility of LRG lensed Einstein rings, it can also be easily applied to Einstein rings / arcs from galaxy cluster lenses, assuming we know the positions of the galaxy clusters from BCG catalogues. In a more general context, such ring finder algorithms have wider applications in finding any circular patterns, for example, planetary nebulae, circumstellar shells around mass loss stars (AGB stars or LBVs), supernovae remnants, to name a few. The disadvantage of finding circular patterns around these astronomical objects, even if their locations can be inferred using certain selection criteria, is the large parameter space of their radii. The unknown radii will make traditional Hough transform computational expensive. To improve on this, there are some variants of circle Hough transform, e.g. probabilistic Hough transform and randomized Hough transform have been proposed. For example, Chiu, Lin, & Liaw (2010) have proposed a fast randomized Hough transform algorithm, which selects a random point on the image as seed point, and using a dedicated checking rule to confirm if the seed point is on a true circle or not. They have shown that such approach greatly reduces the computation effort, and is also suitable for noisy images.

---

[10]https://spacewarps.org/

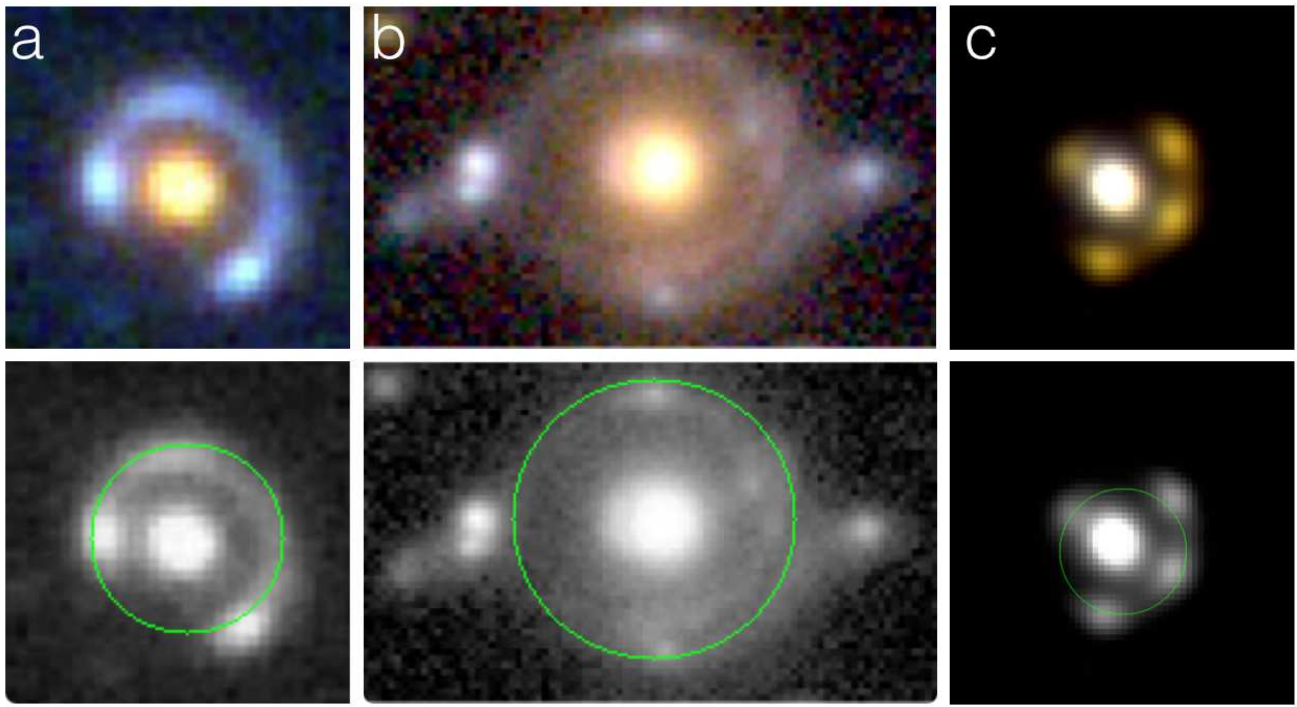

**Figure 2.** Applying Hough transform to Einstein rings from DECam, HSC, and UltraVISTA. Upper panels: images of a) Canarias Einstein ring (from DES Bettinelli et al. 2016); b) Eye of Horus (from HSC Tanaka et al. 2016); c) COSMOS 0050+4901 (from UltraVISTA Hill et al. 2016). Lower panels: Gray-scale images used for patterns recognition, and circular patterns identified by Hough transform (marked in green circles).

In section 3 we mentioned the computational requirements for Hough transform to go through each individual image, but not the total performance time, including pre-processing. In fact, the pre-processing will be the most time-consuming parts, including query of the candiate LRGs to obtain their coordinates, and subsequently making postage stamp images of these LRGs. In the case of SDSS, with its convenient SQL interface, we can easily query and obtain the coordinates of all of the candidate LRGs within a few hours. Given the coordinates, we can obtain the postage stamp in batch mode, which will cost less than an hour. As most of the on-going and forth-coming all sky surveys (e.g. Pan-STARRS, HSC-SSPw) will also provide SQL interface for fast data-base query, as well as postage stamp service that can be fetched in batch mode, we expect that pre-processing time will be in the order of few hours.

In this study we only discuss recognition of circular patterns. Nevertheless, it is possible to apply Hough transform for elliptical patterns. Such algorithms will be very useful in identifying Einstein cross, or quadruply lensed quasars (quads), where one pair of the four lensed quasars have larger projected distance to the lens galaxy than the other pair, and can be recognized as an ellipse. Quads are of great interests of the gravitational lensing community, not only because they can provide more constraints on the lens mass profile, but also due to their potential to determine Hubble constant using time-delay methods. With the advent of wide area sky surveys, Oguri & Marshall (2010) have predicted hundreds lensed quasars in HSC, thousands of lensed quasars in Pan-STARRS and Dark Energy Survey, and ∼ ten thousand of lensed quasars in the up-coming Large Synoptic Survey Telescope, where 10-15% of them will be quads. In such wide area surveys, visual inspection to find quads is extremely difficult, if not impossible. Thus, efficient quads finding algorithm such as elliptical Hough transform will greatly reduce human interactions and provide an automated, robust method to identify interesting quads for the surveys to come.

## 5 ACKNOWLEDGEMENTS

Funding for the Sloan Digital Sky Survey IV has been provided by the Alfred P. Sloan Foundation, the U.S. Department of Energy Office of Science, and the Participating Institutions. SDSS-IV acknowledges support and resources from the Center for High-Performance Computing at the University of Utah. The SDSS web site is www.sdss.org.

SDSS-IV is managed by the Astrophysical Research Consortium for the Participating Institutions of the SDSS Collaboration including the Brazilian Participation Group, the Carnegie Institution for Science, Carnegie Mellon University, the Chilean Participation Group, the French Participation Group, Harvard-Smithsonian Center for Astrophysics, Instituto de Astrofísica de Canarias, The Johns Hopkins University, Kavli Institute for the Physics and Mathematics of the Universe (IPMU) / University of Tokyo, Lawrence Berkeley National Laboratory, Leibniz Institut für Astrophysik Potsdam (AIP), Max-Planck-Institut für Astronomie (MPIA Heidelberg), Max-Planck-Institut für Astrophysik (MPA Garching), Max-Planck-Institut für Extraterrestrische Physik (MPE), National Astronomical Observatory of China, New Mexico State University, New York University, University of Notre Dame, Observatário Nacional / MCTI, The Ohio State University, Pennsylvania State University, Shanghai Astronomical Observatory, United Kingdom Participation Group, Universidad Nacional Autónoma de México, University of Arizona, University of Colorado Boulder, University of Oxford, University of Portsmouth, University of Utah, University of Virginia, University of Washington, University of Wisconsin, Vanderbilt University, and Yale University.